\documentclass[prd,aps,eqsecnum,floatfix,nofootinbib,preprint,tightenlines]{revtex4}
\usepackage{latexsym}
\usepackage{graphicx}
\usepackage{amsmath}
\usepackage{hyperref}
\def\NON{\nonumber\\}
\def\bibi{\bibitem}

\def\a{\alpha}

\def\c{\chi}
\def\d{\delta}
\def\e{\epsilon}                % Also, \varepsilon
\def\f{\phi}                    %       \varphi
\def\g{\gamma}

\def\j{\psi}

\def\l{\lambda}
\def\m{\mu}

\def\p{\pi}                     % Also, \varpi
\def\r{\rho}                    %       \varrho
\def\t{\tau}

\def\x{\xi}

\def\J{\Psi}

\def\ca{{\cal A}}
\def\cb{{\cal B}}

\def\cd{{\cal D}}

\def\cf{{\cal F}}

\def\ch{{\cal H}}   % overridden by cosh !!
\def\ci{{\cal I}}

\def\cm{{\cal M}}

\def\cp{{\cal P}}

\def\cu{{\cal U}}

\def\cbo{{\,\raise-.15ex\Sc [\,}}                       % curly "
\def\sbra#1{\left\langle #1\right|}             % variable < |
\def\sket#1{\left| #1\right\rangle}             % variable | >
\def\svev#1{\left\langle #1\right\rangle}       % variable < >
\def\ddt#1{{\buildrel {\hbox{\LARGE .\kern-2pt.}} \over {#1}}}% double dot-over
\def\ie{\mbox{\it i.e.} }
\def\eg{\mbox{e.g.} }

\def\leqx{\,\raisebox{-1.0ex}{$\stackrel{\textstyle <}{\sim}$}\,}

\def\sbraket#1#2{\left\langle#1|#2\right\rangle} % variable < | >

\def\det{{\rm det}}

\def\bj{\overline{\j}}
\def\bJ{\overline{\J}}

\def\tca{\tilde{\cal A}}

\def\textit#1{{\it #1}\kern.1em }
\def\ie{{\it i.e.}}
\def\cf{{\it cf.}}
\def\eg{{\it e.g.}}

\begin{document}
\hyphenation{fer-mio-nic per-tur-ba-tive pa-ra-me-tri-za-tion
pa-ra-me-tri-zed a-nom-al-ous}

\renewcommand{\thefootnote}{$*$}

\begin{center}
\vspace*{10mm}
{\large\bf The Tunneling Hybrid Monte-Carlo algorithm}
\\[12mm]
Maarten Golterman$^a$\ and\ Yigal Shamir$^b$
\\[8mm]
{\small\it
$^a$Department of Physics and Astronomy,
\\San Francisco State University,
San Francisco, CA 94132, USA}
%\\{\tt maarten@stars.sfsu.edu}
\\[5mm]
{\small\it $^b$School of Physics and Astronomy\\
Raymond and Beverly Sackler Faculty of Exact Sciences\\
Tel-Aviv University, Ramat~Aviv,~69978~Israel}
%\\{\tt shamir@post.tau.ac.il}
\\[10mm]
{ABSTRACT}
\\[2mm]
\end{center}

\begin{quotation}
The hermitian Wilson kernel used in the
construction of the domain-wall and overlap Dirac operators
has exceptionally small eigenvalues that make it  expensive to
reach high-quality chiral symmetry for domain-wall fermions,
or high precision in the case of the overlap operator.
An efficient way of suppressing such eigenmodes consists of including a
positive power of the determinant of the Wilson kernel in the Boltzmann weight,
but doing this also suppresses tunneling between topological sectors.
Here we propose a modification of the Hybrid Monte-Carlo algorithm which
aims to restore tunneling between topological sectors by excluding
the lowest eigenmodes of the Wilson kernel
from the molecular-dynamics evolution,
and correcting for this at the accept/reject step.  We discuss the
implications of this modification for the acceptance rate.

\end{quotation}

\renewcommand{\thefootnote}{\arabic{footnote}} \setcounter{footnote}{0}

%%%%%%%
\newpage
\section{\label{intro} Introduction}
%%%%%%%
The domain-wall-fermion (DWF) \cite{kaplan,ys} formulation of lattice
QCD is based on a hermitian, four-dimensional, super-critical Wilson-like
kernel $H_W=\gamma_5 D_W$, which can be interpreted as governing propagation
into an extra, fifth lattice dimension with coordinate
$s=1,\dots,L_5$.\footnote{
  There is much freedom in choosing the super-critical kernel,
  and the detailed form
  is not important for this paper.  For simplicity, we will assume
  the common choice that $D_W$ is the standard Wilson operator with bare mass
  $m_0$
  in the region $-2 < a m_0 < 0$.  The domain-wall height is $M=|am_0|$.
}
For sufficiently large $L_5$, right-handed and left-handed
four-dimensional quark fields live on walls
opposite of each other across this fifth dimension, and are represented
by wave functions which are bound to these walls.  As long as $L_5$ is
finite, the right-handed and left-handed
quarks are coupled through a ``residual'' mass
\cite{mresCPPACS,mresRBC,yspt}, an effective
chiral-symmetry breaking mass term originating
from the overlap of their wave functions.

In the free theory \cite{kaplan,pvsm} or in perturbation theory \cite{AT,yspt}
these wave functions are exponentially bound to the walls, and therefore
the residual mass is exponentially small.  It is proportional to
$e^{-\a_{min} L_5}$ where the gap of the transfer-matrix hamiltonian,
$\a_{min}$, is non-zero whenever the gap of the free $H_W$ is.
On realistic gauge field configurations, however, the super-critical
$H_W$ has a spectrum reaching all the way down to zero \cite{scri,bnn},
with a mobility edge that, by definition, separates a low-eigenvalue
spectrum of localized modes from a high-eigenvalue spectrum of
delocalized, or extended, modes.  The near-zero modes
are localized with a range of order the lattice spacing $a$, provided that
the bare gauge coupling $g$ and domain-wall height $M$ are chosen to be away
from the region in the phase diagram  where the mobility edge is less than
order one in lattice units \cite{gs}.

An additional contribution to the residual mass
is coming from the near-zero modes.  It can be approximated by
the integrated spectral density $\int_{-1/L_5}^{1/L_5} d\l\, \r(\l)$,
where $\rho(\l)$ is the ensemble-average spectral density of $H_W$
(see Appendix~C of Ref.~\cite{yspt}).  If the near-zero spectral density
depends only mildly on $\l$, this contribution is roughly equal
to $\r(0)/L_5$.  Therefore, as $L_5$ is increased, eventually
the slowly falling contribution of the near-zero modes
becomes dominant. Much effort has been directed toward
finding methods to suppress their presence, \ie, to make $\rho(\l)$ as
small as possible for $|\l| \leqx 1/L_5$.
The same goal is relevant for  overlap simulations,
where suppression of near-zero modes is equally important in order to
make it affordable to compute the sign function $\e(H_W)=H_W/|H_W|$,
which is at the heart of the construction of the overlap operator \cite{hn},
to high precision.

A ``surgical'' method for suppressing the near-zero modes consists of
modifying $\cb$, the original domain-wall or overlap Boltzmann weight,
to $\cb\, \det(H_W^2)$ \cite{pv,jlqcd}.
The first thing to note is that
introducing the auxiliary determinant of $H_W^2$
into the Boltzmann weight
does not change
the universality class. Indeed, apart from a change in the lattice spacing,
this modification does not affect the long-distance physics,
because in itself the Wilson kernel $H_W$
is a Dirac operator with a mass $m_0$ of the
order of the cutoff, $|am_0|=O(1)$.

It has been demonstrated that, after re-adjusting the lattice spacing to
the same value, the inclusion of $\det(H_W^2)$ in the path integral
results in a significant reduction in the near-zero mode density
\cite{pv,jlqcd}. This is easily understood at an intuitive level because
the Boltzmann weight now contains an extra factor of $\l^2$
for each eigenmode of $H_W$, including in particular
the low-lying, localized ones.
Provided we stay
away from the region where the mobility edge is too small,
the near-zero modes of $H_W$ are associated with
lattice-scale dislocations in the gauge field \cite{bnn}.
One may expect
the inclusion of $\det(H_W^2)$ to deplete the near-zero modes efficiently
because this method suppresses selectively
only those---relatively rare---dislocations that happen
to support a near-zero mode.  In other words, once we re-adjust
the bare coupling such that the gauge field regains the same
average ``roughness'' as before the insertion of $\det(H_W^2)$,
and hence the lattice spacing is unchanged, we find that
only the ``troublesome'' dislocations leading to exceptionally small
eigenmodes of $H_W^2$ have been removed.
The new spectral density behaves like
$\r(\l)\sim \l^2$ at small $\l$, and the contribution to the residual mass is
of order $\int_{-1/L_5}^{1/L_5} d\l\; \r(\l)\sim 1/L_5^3$,
to be compared with a $1/L_5$ suppression
for a spectral density that does not vanish at $\l=0$.

In spite of their largely negative role discussed above,
the near-zero modes also play what might be called a positive role.
The gauge-field configuration space is divided into topological sectors
whose boundaries are defined by configurations where $\det(H_W)=0$.
If we continuously follow a curve in the gauge-field space,
it follows that the topological
charge, as measured through the index of the overlap operator, changes
by one unit precisely when one of the eigenvalues of $H_W$ changes
sign \cite{topo}.

To date, the Hybrid Monte Carlo (HMC) algorithm \cite{hmc}
is the \textit{de facto} algorithm for dynamical fermion simulations.
The question is whether the trajectories generated by the HMC
algorithm, or by one of its variants, will sample all
topological sectors.
Having fewer, even much fewer, near-zero eigenvalues would not
constitute a problem all by itself, so long as these eigenvalues could move
around and cross zero relatively freely.
Unfortunately, the inclusion of $\det(H_W^2)$ not only reduces the
near-zero spectral density significantly, but also completely suppresses
the transitions between topological sectors.  The reason is that,
during the molecular-dynamics (MD) evolution phase
of the HMC algorithm,  the  surfaces of co-dimension one
where $\det(H_W)=0$ constitute infinite-energy barriers. As a result,
only one topological sector is sampled, typically
the sector with zero topological charge.
This constitutes a breakdown of ergodicity.
In other words, the effect of adding in $\det(H_W^2)$ while
using the standard HMC algorithm is to generate
an ensemble according to the Boltzmann weight $\cb\, \det(H_W^2) \d(Q-q)$,
where $Q$ is the topological charge operator, and $q$ the topological
charge of the initial configuration.
Thus the correct weight $\cb\, \det(H_W^2)$
is reproduced in just one topological sector,
while the weight vanishes in all other sectors.\footnote{
  We assume that there are no exactly massless quarks, and thus
  $\cb>0$ in all topological sectors.
}

Unlike $\theta$-vacua, fixed-topological-charge vacua do not cluster.
This leads to enhanced finite-volume effects,
which, already for single-particle masses,
decrease only with an inverse power of the volume,
and, moreover, increase with the inverse pion mass \cite{bcnw}.
This must be compared with the usual exponential fall-off of finite-size
effects in QCD with massive quarks.

In this paper, we propose a modification of the HMC algorithm
intended to restore ergodicity by allowing for zero crossings
in the spectrum of $H_W$ even though $\det(H_W^2)$ is part
of the Boltzmann weight.  The generated ensemble will then reproduce the
correct Boltzmann weight  $\cb\, \det(H_W^2)$
for configurations in all sectors.
We will refer to this modification as the ``Tunneling Hybrid
Monte Carlo'' or THMC algorithm.

The key idea is that, during the MD
evolution, the usual HMC fermion force associated with $H_W^2$
is replaced by the force derived from
a modified operator of the form of $H_W^2+\a Q^\dagger Q$.  The effect of
the new term, $\a Q^\dagger Q$, is to lift  the (lowest few) near-zero
eigenvalues of $H_W^2$.  This removes the infinite-energy barriers.
As a result, the MD evolution can, and, we argue, will, visit all
topological sectors.
Using a trick reminiscent of renormalization-group blocking,
the original determinant $\det(H_W^2)$ can be factorized
as $\det(H_W^2+\a Q^\dagger Q)$ times another term that corrects for
the lifted near-zero modes.  The original Boltzmann weight
$\cb\,\det(H_W^2)$ is restored
by incorporating the correcting factor
in the accept/reject step that separates successive MD trajectories.
While this procedure must lead to some
reduction in acceptance, we will give arguments suggesting
that that reduction may be affordable,
given the physical merits of the generated ensemble.
This paper does not as yet contain any numerical tests of the algorithms;
these will have to be carried out in the future.

This paper is organized as follows.  In Sec.~\ref{basic}
we set up the framework
and explain the basic idea.  In Sec.~\ref{determine}
we discuss the conceptually simplest
realization of the algorithm, in which the new term that lifts
the near-zero eigenvalues, $\a Q^\dagger Q$, is chosen deterministically.
The resulting implementation of the algorithm may not be very practical
because of its cost.  A more affordable
implementation of the algorithm, introduced in Sec.~\ref{stoch},
is based on a stochastic choice of the new term.
In Sec.~\ref{par} we discuss reasonable choices for the parameters
of the algorithm, and we end with our conclusions in Sec.~\ref{disc}.
Some technicalities are relegated to two appendices.

%%%%%%%
\section{\label{basic} The basic idea}
%%%%%%%
We will consider a lattice gauge theory with partition function
\begin{subequations}
\label{Z}
\begin{eqnarray}
Z&=&\int\cd\cu\;\exp(-S_g(\cu))\;\det(H_W^2(\cu))
\label{Za}\\
&=&\int\cd\cu\cd\phi^\dagger
\cd\phi\;\exp(-S_g(\cu)-S_{pf}(\cu))\ ,
\label{Zb}\\
S_{pf}&=&\phi^\dagger H_W^{-2}(\cu) \phi\ ,
\label{Zc}
\end{eqnarray}
\end{subequations}
where $S_g$ is the (unspecified) gauge action, and once again $H_W$ is the
hermitian, super-critical Wilson operator. The QCD partition
function should of course also include the physical domain-wall or overlap
fermion determinant for each quark flavor.
However, since we are concerned only with the effects of including
the unphysical determinant $\det(H_W^2)$, we have dropped the
determinants representing the physical quarks.
The continuum limit of the lattice theory~(\ref{Z}) is thus a pure
Yang-Mills theory.

In Eq.~(\ref{Zb}) the determinant of $H_W^2$ has been rewritten as a pseudo-fermion
partition function. If ergodicity were to hold,
the HMC algorithm would generate an ensemble of
configurations $\cu_i$ with a probability measure
\begin{equation}
\label{prob}
\cp_g(\cu)=Z^{-1}\;\exp(-S_g(\cu))\;\det(H_W^2(\cu))\ ,
\end{equation}
by alternately updating the pseudo-fermion field $\f$ and updating the gauge
field $\cu$.  As explained in the Introduction,
the standard HMC algorithm in fact fails to be ergodic
because of the zero modes of $H_W$, and this is what we seek to amend
with the THMC variant.

Let us recall how the HMC algorithm normally works
(for recent reviews, see Ref.~\cite{rev}).  An updating cycle
begins by generating a new pseudo-fermion field
via a heat bath.  A random complex vector $\x$ is drawn
from a gaussian ensemble with unit width, that is, with
probability distribution
$\cp(\x) \propto \exp(-\x^\dagger \x)$, followed by setting
\begin{equation}
\label{phix}
  \phi=H_W(\cu)\x\ .
\end{equation}
Thus the probability distribution for $\f$ is
$\cp(\f) \propto \exp(-\f^\dagger H_W^{-2}(\cu) \f)$.
A set of fictitious momenta is similarly drawn,
corresponding to the free action
$S_\pi=\frac{1}{2}\sum_{x,\mu,a}\pi_{x,\mu,a}^2$.
The gauge-field update then consists of two steps.
First, the initial configuration of gauge field and conjugate momenta,
$\{\cu,\p\}$, is evolved along
an MD trajectory by numerically solving the classical Hamilton equations
with some ``guiding'' hamiltonian $\ch_{MD}$.
In the most straightforward case, one takes as
guiding hamiltonian
\begin{equation}
\label{HMD}
\ch_{MD}=S_\pi(\pi)+S_g(\cu)+S_{pf}(\cu)\ .
\end{equation}
The gauge-field dependence of $S_{pf}$ comes from
the operator $H_W^{-2}(\cu)$ in Eq.~(\ref{Zc}).
We have suppressed the dependence of $S_{pf}$ on the
pseudo-fermion field $\phi$, which is kept fixed during the MD evolution.
Hamilton's equations are numerically solved
with some symmetric integrator with a non-zero step size
$\delta\tau$, for a number of steps $n_{MD}=\tau_{MD}/\delta\tau$.  This
introduces ``step-size'' errors, which at the end of an MD trajectory
are corrected for by a Metropolis test: the final configuration $\{\cu',\p'\}$
is accepted, which means that $\cu'$ becomes the initial gauge-field
configuration for the next cycle, or rejected,
in which case the next cycle begins
with the same initial gauge-field configuration $\cu$, with probability
\begin{equation}
\label{ar}
P_{accept}={\rm min}\{1,\exp(\ch(\pi,\cu)-\ch(\pi',\cu'))\}\ ,
\end{equation}
where we have taken $\ch=\ch_{MD}$.
Assuming ergodicity, after thermalization the generated sets
of configurations $\{\cu_i,\p_i,\f_i\}$
follow the probability distribution
\begin{equation}
\label{Pall}
\cp(\cu,\p,\f)=Z^{-1}\;\exp(-\ch(\cu,\p,\f))\ ,
\end{equation}
which, upon integrating over the pseudo-fermions and fictitious momenta,
reduces to the probability distribution (\ref{prob}).

In many applications, the randomness of the fictitious momenta and
the pseudo-fermions at the start of each MD trajectory
would be sufficient to make the algorithm ergodic \cite{dk}.
However, in the case at hand this is hampered
by the existence of zero modes of $H_W$. During the MD evolution
$\ch_{MD}(\cu,\p,\f)$ plays the role of ``energy.''
The pseudo-fermion action, which is part of $\ch_{MD}(\cu,\p,\f)$,
becomes infinite whenever $H_W(\cu)$ has an exact zero mode.
Because the initial energy is always finite, and since it is (approximately)
conserved during the (discretized) MD evolution, (in practice) the MD evolution
never passes through any configuration where $\det(H_W(\cu))=0$.\footnote{
  We note that it is not possible to regain tunneling
  by choosing $\d\t$ to be very large because this must lead to significant
  drop in acceptance and ultimately to integrator instabilities \cite{rev}.
}
As explained in the introduction, this implies that only one topological
sector is sampled.

What one would like to do is to somehow remove the infinite-energy barriers
separating topological sectors during the MD evolution.
The basic philosophy of the HMC algorithm allows for this option:
one is free to choose $\ch$ in Eq.~(\ref{ar}) different from the guiding hamiltonian
$\ch_{MD}$. Eliminating the  infinite-energy barriers necessarily means
that $\ch_{MD}$ cannot be given by
Eq.~(\ref{HMD}) if $S_{pf}$ is given by Eq.~(\ref{Zc}). But any ``mistake'' made by
choosing a different  guiding hamiltonian can be
corrected for at the accept/reject step if the appropriate hamiltonian
is used in Eq.~(\ref{ar}).  It should of course be anticipated that
the discrepancy between the two hamiltonians will lead to reduced acceptance,
an issue that we will return to below.

We thus wish to split the hamiltonian of the metropolis test into two parts:
\begin{equation}
  \ch=\ch_{MD}+\ch_{corr}\ ,
\label{Hcorr}
\end{equation}
where contributions of near-zero modes
are kept out of the guiding hamiltonian $\ch_{MD}$, and are accounted
for by the correction $\ch_{corr}$, to be added
in the accept/reject step.  This can be done as follows.  Borrowing from
the way ultra-violet modes are integrated out in a Renormalization-Group
(RG) block transformation for a bilinear action, one can
show that the fermion determinant can be split as
\begin{subequations}
\label{RG}
\begin{eqnarray}
\det(H_W^2) &=& \a^{-n}\; \det(\cm)\;\det(\ca)\ ,\label{RGa}\\
\cm &=& H_W^2+\a Q^\dagger Q\ ,\label{RGb}\\
\ca^{-1}&=&\a^{-1}+Q \, H_W^{-2} \, Q^\dagger\label{RGc}\ .
\end{eqnarray}
\end{subequations}
Here $H_W$ and $\cm$ are $N\times N$ matrices, $\ca$ is an $n\times n$ matrix,
and $Q$ is an $n\times N$ matrix, usually referred to as the blocking kernel
in the RG context. For a proof of Eq.~(\ref{RG}), see App.~\ref{appA}.\footnote{
  Equation (\ref{RG}) is true for
  an arbitrary choice of the kernel matrix $Q$ and of the parameter $\a\ne 0$.
  There is no restriction on $n$ and $N$
  (curiously, it is valid even for $n>N$).
  Furthermore, Eq.~(\ref{RG}) holds in a limiting sense even if $H_W$ has an
  exact zero mode.
}

Here we will choose the kernel $Q$ such that $Q^\dagger Q$ projects
onto the lowest $n$ eigenmodes of $H_W^2$,
either exactly or approximately.\footnote{
  As discussed later in detail, we have in mind choosing $n$ very small
  and, thus, $n\ll N$.
}
Let $|\psi_i\rangle$, $i=1,\dots,n,$
be the eigenvectors of the first-order operator $H_W$ with
eigenvalues $\l_i$ chosen such that $\{\l_i^2\}_{i=1}^n$ is the set
of $n$ smallest eigenvalues of $H_W^2$.
Now choose another set of $n$
vectors $|\chi_i\rangle$ such that
\begin{equation}
\label{I}
\ci_{ij}=\langle\psi_i|\chi_j\rangle\approx\delta_{ij}\ ,
\qquad \qquad i,j=1,\ldots,n\ ,
\end{equation}
and define $Q^\dagger$ to be the $N\times n$ matrix whose $i$-th column is
the vector $|\chi_i\rangle$.
Equations~(\ref{RGb}) and (\ref{RGc}) then turn into
\begin{subequations}
\label{MA}
\begin{eqnarray}
\cm &=& H_W^2 + \a\sum_{i=1}^n|\chi_i\rangle\langle\chi_i|\ ,\label{MA1}\\
(\ca^{-1})_{ij}&=&\a^{-1}\d_{ij}
+\langle\chi_i| H_W^{-2} |\chi_j\rangle\ .\label{MA2}
\end{eqnarray}
\end{subequations}
Assuming Eq.~(\ref{I}), the rightmost term in Eq.~(\ref{MA1}) lifts
the $n$ lowest eigenvalues of $H_W^2$, and the corresponding eigenvalues
of $\cm$ are of order $\a$.  All the remaining eigenvalues of $H_W^2$
are more or less unaffected, if we choose
$\a$ to be of order $\l_{n+1}^2$.

Using Eq.~(\ref{RGa}) we may write\footnote{
  Following common practice to leave unspecified the overall normalization of
  a path-integral measure, we disregard from now on
  the normalization factor $\a^{-n}$ in Eq.~(\ref{RGa}).
}
\begin{subequations}
\label{pf}
\begin{eqnarray}
  \det(H_W^2) &=& \det(\ca) \int\cd\phi^\dagger\cd\phi\;\exp(-S_{pf})\ ,
\label{pfa}\\
  S_{pf} &=& \phi^\dagger \cm^{-1} \phi\ .
\label{pfb}
\end{eqnarray}
\end{subequations}
The THMC algorithm is based on choosing $\ch_{MD}$ as in Eq.~(\ref{HMD})
with $S_{pf}$ given by Eq.~(\ref{pfb}).  In order to recover the desired
Boltzmann weight (\ref{prob}),  we will include $\det(\ca)$
in the correction term $\exp(-\ch_{corr})$, \cf~Eqs.~(\ref{Hcorr}) and (\ref{ar}).
The definition of the THMC algorithm is not complete until
a method for choosing the vectors $\sket{\c_i}$ is specified.
There are two basic options: the $|\chi_i\rangle$ can be chosen deterministically
or stochastically, and this
leads to the concrete implementations of THMC presented
in Sec.~\ref{determine} and Sec.~\ref{stoch}, respectively.   In the latter case,
$\exp(-\ch_{corr})$ is not precisely
equal to $\det(\ca)$, as we will see.
In both implementations the gauge field will be distributed according to
the probability measure (\ref{prob}).

Using only the new form
of the guiding hamiltonian, it is easy to see that all topological
sectors will now be sampled.
Any configuration at a crossing point
between two adjacent topological sectors has
an exact zero mode of $H_W$.  But, so long as Eq.~(\ref{I}) holds,
Eq.~(\ref{MA1}) implies that the spectrum of $\cm$ has an $O(\a)$ gap,
even if $H_W$ happened to have an exact zero mode.
Thus, the excess MD energy associated with a crossing is of order
$1/\a$.
This is a finite number, because of our choice of $\a$.
This means that the initial MD energy has a considerable probability
of being larger than the minimal MD energy needed for the crossing.
As a result, the MD trajectory can pass through configurations
where $H_W$ has an exact zero mode.  Typically, both at the start and end
points of an MD trajectory the gauge-field configuration will be well inside
some topological sector, so that neither $\det(\ca(\cu))$ nor $\det(\ca(\cu'))$
will be exceedingly small. This already suggests that there is no inherent
reason that the acceptance rate should be intolerably small.
We will discuss the issue of acceptance when introducing concrete
implementations in the next two sections.

%%%%%%%
%\newpage
\section{\label{determine} Deterministic implementation}
%%%%%%%
In this section we present the conceptually simplest implementation
of THMC.  It is defined by setting
\begin{equation}
\label{oi}
|\chi_i\rangle=|\psi_i\rangle\ ,
\end{equation}
at each time step during the MD evolution, where $\sket{\j_i}$ are
the $n$ lowest eigenmodes of $H_W^2$ on the updated gauge field.
It follows that $\ci_{ij}=\delta_{ij}$ in Eq.~(\ref{I}), and also that
\begin{subequations}
\label{AM}
\begin{eqnarray}
  \cm &=& H_W^2 + \a\sum_{i=1}^n |\psi_i\rangle\langle\psi_i| \ ,
\label{AMa}
\\
  (\ca^{-1})_{ij} &=& \rule{0ex}{3ex}
  \Big(\a^{-1}+\l_i^{-2}\Big)\d_{ij}\ .
\label{AMb}
\end{eqnarray}
\end{subequations}
The representation of the partition function which is being simulated is thus
\begin{equation}
\label{Zov}
  Z = \int\cd\cu\cd\phi^\dagger\cd\phi\;
  \det(\ca)\,\exp(-S_g-S_{pf})\ ,
\end{equation}
with $S_{pf}$ given by Eq.~(\ref{pfb}).
The guiding hamiltonian is given by Eq.~(\ref{HMD}), and
the accept/reject step is done with
\begin{equation}
\label{ARmod}
P_{accept}={\rm min}\left\{1,\exp(\ch_{MD}(\pi,\cu)-\ch_{MD}(\pi',\cu'))\;
\frac{\det(\ca(\cu'))}{\det(\ca(\cu))}\right\}\ .
\end{equation}
Comparing to Eqs.~(\ref{ar}) and (\ref{Hcorr}), we see that this
amounts to choosing $\ch_{corr}=-\log\det(\ca)$.  We note that
since $\cm$ cannot be written in the form $X^\dagger X$ with
$X$ local, one updates the pseudo-fermions via $\phi=r(\cm(\cu))\x$
with some high-accuracy rational approximation\footnote{
  By construction, $\cm$ does not have any extremely small eigenvalues,
  and we expect that an accurate rational approximation of
  $\sqrt{\cm}$ is feasible with relatively few poles.
}
$r(x)\simeq\sqrt{x}$.

While conceptually the simplest, the deterministic implementation of THMC
involves several costly ingredients.
The $n$ smallest eigenvalues of $H_W^2$
and the corresponding eigenmodes must be re-calculated
after each MD step.
For the calculation of the fermion force we have to monitor
for level crossings during each MD time increment.\footnote{
  We assume that for some $n'>n$, the $n'$ lowest eigenmodes evolve
  slowly enough that one can track their evolution
  from time $\t$ to $\t+\d\t$.
}
If no level crossings occurred,
the calculation of the fermion force requires
\begin{subequations}
\label{dpsi}
\begin{eqnarray}
\label{ff}
\d_{x,\m}\left(|\psi_i\rangle\langle\psi_i |\right)
&=&(\d_{x,\m}|\psi_i\rangle)\langle\psi_i |+
|\psi_i\rangle(\d_{x,\m}\langle\psi_i |)\ ,
\label{dpsia}
\\
\d_{x,\m}|\psi_i\rangle &=& \rule{0ex}{2.5ex}
-\left(H_W-\l_i\right)^{-1}
\left(1-|\psi_i\rangle\langle\psi_i |\right)
\left(\d_{x,\m} H_W \right)|\psi_i\rangle\ ,
\label{dpsib}
\end{eqnarray}
where the last equality follows from first-order perturbation theory, and
\begin{equation}
\label{delta}
\d_{x,\m}=i\left(U_{x,\m}\;
\frac{\partial{~~~}}{\partial U^T_{x,\m}}-{\rm h.c.}\right)\ .
\end{equation}
\end{subequations}
The numerical evaluation of $\d_{x,\m}|\psi_i\rangle$ using Eq.~(\ref{dpsia})
may be difficult in the vicinity of a level-crossing point.
For a practical solution to this problem, see \eg~Ref.~\cite{sdg}.\footnote{
  In an overlap simulation, it is known that
  the presence of roughly degenerate near-zero eigenvalues of $H_W$
  may lead to large fermion forces and to low acceptance rates.
  Using THMC may help indirectly, because the presence of $\det(H_W^2)$
  will lead to much reduced density of low-lying eigenvalues.
  For a direct approach to solving this problem, see Ref.~\cite{NC}.
}

The situation is different if level crossing occurs between
levels $n$ and $n+1$.  In this case the set of $n$ lowest eigenmodes
changes discontinuously which, according to Eq.~(\ref{AMa}), results
in a discontinuity in the operator $\cm$.
The discontinuous change in $S_{pf}$ and, by Eq.~(\ref{HMD}), in
the MD energy, gives rise to a $\d$-function singularity in the fermion force.
This singularity must be handled by a reflection/refraction step
analogous to that discussed in Ref.~\cite{rflr}.  The discussion
of Ref.~\cite{rflr} addresses specifically the discontinuity
of the overlap operator arising from a zero crossing in the spectrum of
the corresponding kernel operator, here assumed to be $H_W$.
But the recipe of Ref.~\cite{rflr} is actually rather general,
and can be easily adapted to the case at hand.
This completes the definition of
the deterministic implementation of THMC.

We comment in passing that one might be tempted to consider the
alternative of evolving continuously the $n$ eigenmodes
that were the lowest levels on the initial configuration.
This, however, would constitute a breakdown of reversibility of
the MD evolution whenever the set of $n$ lowest eigenmodes
on the final configuration is different from the $n$ states
that have evolved continuously from the initial set.\footnote{
  We thank Urs Heller for this observation.
}
Thus, reflection/refraction steps are unavoidable.

The cost of the deterministic implementation of THMC involves two
components: first, the extra cost of each MD step, and second, the reduction in
acceptance due to the omission of $\det(\ca)$ from the MD evolution.
We discuss them in turn.

In a dynamical overlap simulation,\footnote{
  For a recent review, see Ref.~\cite{sf}.
}
in order to improve the accuracy of the evaluation
of the sign function of $H_W$ one usually computes $n_{ov}$ of the
lowest eigenvalues of $H_W^2$, alongside with their eigenfunctions and the
variations defined in Eq.~(\ref{dpsi}).  Thus, so long as we choose
$n$ in Eq.~(\ref{MA}) to be less than or equal to $n_{ov}$, there is no
extra cost involved in using them for the deterministic THMC algorithm.

The exception is the occurrence of level crossing between eigenvalues $n$
and $n+1$ of $H_W^2$ which, as explained above, must be handled by a
reflection/refraction step.  This new reflection/refraction step
represents an additional cost not present in an ordinary dynamical overlap
simulation.  We stress that the discontinuity encountered
here results directly from the breakup of $\det(H_W^2)$ in Eq.~(\ref{RGa}).
The discontinuity in $\det(\cm)$ is matched by a discontinuity
in $\det(\ca)$,  such that $\det(H_W^2)$ as a whole is continuous.
Also, clearly, in this case there is no discontinuity in
the corresponding overlap operator, because no eigenvalue of $H_W$ has
crossed zero.
The cost of this new reflection/refraction step could be more tolerable
than in the familiar overlap case, because
the eigenfunctions of the crossing eigenvalues are both
expected to be well localized, hence the cost of the necessary computation
is likely to be independent of the volume.  In addition,
also the number $n$ of eigenmodes kept out of the MD evolution
is held fixed and small, independent of the volume (see also Sec.~\ref{disc}).

Omitting $\det(\ca)$ from the MD evolution, or equivalently,
performing the metropolis test with a hamiltonian $\ch$
which is different from the guiding hamiltonian $\ch_{MD}$, will in general
decrease the acceptance.  In App.~\ref{acc} we
consider a crude model to estimate this effect.
According to this model,
when a single eigenmode is omitted from the MD evolution,
the drop in acceptance is expected to be in the range of $1/3$ to,
at most, $1/2$.  When two eigenmodes are omitted, the drop in acceptance
is in the range of roughly 50\% to 65\%.  It is thus desirable
to keep the number of eigenmodes relegated to  $\det(\ca)$ as small as
possible, perhaps even $n=1$.  We return to this issue in Sec.~\ref{par}.
In order to avoid confusion,
we stress that in the stochastic implementation discussed in the next section
there is an additional source for a reduction in acceptance.

%%%%%%%
\section{\label{stoch} Stochastic implementation}
%%%%%%%

We now turn to a different version of THMC in
which the  vectors $\sket{\c_i}$ are kept fixed during the MD evolution.
As we will see, a valid algorithm exists provided that
the  vectors $\sket{\c_i}$ are chosen stochastically.
While the $n$ lowest eigenmodes of $H_W^2$ will have to be computed
at the beginning and the end of each trajectory, this implementation
avoids the costly operations of re-computing these eigenmodes
and their gauge-field derivatives at every MD step.
Also, no reflection/refraction steps are needed for the trivial reason
that the vectors $|\chi_i\rangle$ are not to be evolved.\footnote{
  Of course, in an overlap simulation
  a reflection/refraction step is in principle still needed
  at a zero crossing of an eigenvalue of the kernel \cite{sf}.
}
The stochastic implementation is thus more suitable for domain-wall
fermion simulations, where these calculations are normally not done.

Before we proceed, we need to address a technical issue.
Eigenfunctions of $H_W$ are determined only up to an arbitrary overall phase.
In Sec.~\ref{determine}, the undetermined overall phase dropped out
(see in particular Eq.~(\ref{AM})), and this ambiguity thus was of no concern.
For the stochastic implementation introduced below,
the phase ambiguity will have to be resolved.
If $|\hat\j_i\rangle$ is
an eigenmode with some arbitrary overall phase, we may for example choose
\begin{equation}
  \sket{\j_i} = \frac{\langle\hat\j_i|{\bf 1}\rangle}
                     {|\langle\hat\j_i|{\bf 1}\rangle|}\;
                |\hat\j_i\rangle\ .
\label{fixphase}
\end{equation}
Here $\sket{\bf 1}$ stands for the vector with all component
equal to one.\footnote{
  Other choices for the reference vector can be made in case that
  the inner product $|\langle\hat\j_i|{\bf 1}\rangle|$ would happen to be
  numerically unstable.
}
It is easily seen that $\sket{\j_i}$ is still a normalized
eigenvector, and that it is unchanged if we multiply
$|\hat\j_i\rangle$ by some arbitrary phase.

We are now ready to introduce the stochastic implementation of THMC.
At the start of an MD trajectory we compute, as before, the
lowest $n$ eigenvectors $|\psi_i(\cu)\rangle$,
which are now free of any phase ambiguity by virtue of Eq.~(\ref{fixphase}).
We then draw $n$ random vectors $|\eta_i\rangle$
from a gaussian ensemble with probability
$\exp(-\g{\overline{\eta}}\eta)$, and set
\begin{equation}
\label{chi}
|\chi_i\rangle = |\psi_i(\cu)\rangle + |\eta_i\rangle \ .
\end{equation}
Equivalently, the vectors $\sket{\c_i}$ can be thought of
as $n$ new complex bosonic fields (each carrying the same set of indices as a
Wilson-fermion field) with their own Boltzmann weight $\exp(-S_{ker})$, with
kernel action
\begin{equation}
\label{Sker}
S_{ker}=\g\sum_{i=1}^n\;\langle\chi_i-\psi_i(\cu)|\chi_i-\psi_i(\cu)\rangle\ .
\end{equation}
With the proper (gauge-field-independent) normalization we then have
\begin{equation}
\label{ZQ}
\int\cd\chi^\dagger\cd\chi\;\exp(-S_{ker})=1\ ,
\end{equation}
which we may thus insert into the original partition function Eq.~(\ref{Z}),
obtaining (with Eqs.~(\ref{RGa}) and (\ref{pf}))
\begin{equation}
\label{Zmod}
Z=\int\cd\cu\cd\phi^\dagger\cd\phi\cd\chi^\dagger\cd\chi\;
\exp(-S_g-S_{pf}-S_{ker})\;\det(\ca)\ .
\end{equation}

The relation between the representation (\ref{Zmod}) of the partition function
and the stochastic implementation of THMC parallels the relation
between the representation (\ref{Zov}) and the deterministic implementation.
At the beginning of each trajectory a new set of pseudo-fermions $\f$,
vectors $\sket{\c_i}$, and fictitious momenta $\p$ are drawn, each
from the corresponding heat bath.  The MD evolution is then carried
out with the same guiding hamiltonian as before, \cf~Eq.~(\ref{HMD}).
Finally, the metropolis test (\ref{ar}) is done with the hamiltonian
\begin{equation}
\label{DWFP}
\ch=\ch_{MD}+S_{ker}-\log\det(\ca)\ .
\end{equation}
Note that both $\ch_{MD}$ and $\det(\ca)$ are now functions of the (fixed)
vectors $\sket{\c_i}$, but not directly of any eigenmode of $H_W^2$.
The initial and the final lowest $n$ eigenmodes, $\sket{\j_i(\cu)}$
and $\sket{\j_i(\cu')}$,
are only required for the evaluation of $S_{ker}$.

The stochastic implementation of THMC generates
the correct equilibrium distribution (\ref{prob}), for the same reasons
that the original HMC algorithm works \cite{hmc}.  The algorithmic
role of the new stochastic degrees of freedom, $\sket{\c_i}$,
parallels the role of the pseudo-fermions $\f$ in the standard HMC algorithm
(or, for that matter, in the deterministic implementation of THMC).
In particular, satisfying the detailed balance condition \cite{hmc}
requires that the classical MD evolution be reversible.
This means that if $\{\cu,\p\}$ is evolved into $\{\cu',\p'\}$
then $\{\cu',-\p'\}$ is evolved into $\{\cu,-\p\}$,
\textit{when all other degrees of freedom are held fixed}.
This now includes both $\f$ and $\sket{\c_i}$.
Just like in ordinary HMC, using a symmetric integrator ensures the
reversibility of the MD evolution in the stochastic implementation of THMC.

According to Eq.~(\ref{chi}) the vectors $\sket{\c_i}$ drawn from the heat bath
depend on the initial gauge field.  One might infer
from this equation the erroneous conclusion that ``reversibility
is not satisfied.''  If this were true,
reversibility would not be satisfied in the original HMC algorithm
in the first place,
because according to Eq.~(\ref{phix}) the pseudo-fermion field drawn
from the heat bath depends on the initial gauge field as well.
What drives this confusion is that,
as a whole, the HMC algorithm is not reversible;
it generates a Markov chain that converges
to a specified probability distribution
irrespective of initial conditions, hence irreversibly \cite{dk,hmc,rev}.
The same goes for both versions of THMC.
Reversibility is only required for the classical MD evolution,
where, as stated above, it amounts
to the interchangeability of $\{\cu,\p\}$ and $\{\cu',-\p'\}$
for any given, fixed set of values of the remaining degrees of freedom.
Thus, reversibility of the MD evolution is guaranteed by the use
of a symmetric integrator in the two versions of THMC,
just like in standard HMC.

%%%%%%%
%\newpage
\section{\label{par} Choice of parameters}
%%%%%%%
The THMC algorithm has not been tested yet, and at present it is not
known how well it works in practice.
Assuming some particular choice of the
physical-sector action (including the gauge action and some action
(domain-wall or overlap) for
the physical quarks) there are five parameters that determine
the performance of the stochastic implementation of the
algorithm: $n$, $\a$, $\g$, $\tau_{MD}$ and $\d\tau$.
In Sec.~\ref{basic} we already argued that the ``blocking'' parameter $\a$
should be chosen equal to a typical small eigenvalue of $H_W^2$.
In this section we discuss the remaining parameters of the stochastic
implementation.

We begin with the parameter $\g$
occurring in $S_{ker}$, and explain why this parameter
should be chosen to be of order one.
We will assume that the initial and final low eigenmodes are all
localized and, thus, that their components are of order one within
their region of support. The norm\footnote{
  Because $\sket{\eta}$ is not localized, the magnitude of
  its individual components is of order $1/\sqrt{\g V}$,
  where $V$ is the number of lattice sites.
}
of the random vector $\sket{\eta}$ is of order $1/\sqrt{\g}$.
We immediately see that one cannot choose $\g$
too small, if we want $\sket{\chi_i}$ to be
a reasonable approximation of $\sket{\psi_i(\cu)}$.
If not, there would be no reason for the $n$ lowest eigenvalues of $H_W^2$
to be lifted by the addition of $\a\sum_i\sket{\chi_i}\sbra{\chi_i}$,
\cf~Eq.~(\ref{MA1}).

The parameter $\g$ cannot be chosen arbitrarily large either.
In fact, in the limit $\g\to\infty$ acceptance will vanish.
For $\g\gg 1$ the vectors $\sket{\c_i}$ drawn from the heat bath
at the beginning of the trajectory
will satisfy $\sket{\c_i}=\sket{\j_i(\cu)}+O(1/\sqrt{\g})$.
At the end of the trajectory we will thus have
\begin{equation}
\label{endMD}
  \g\sbraket{\c_i-\j_i(\cu')}{\c_i-\j_i(\cu')}
  =\g\sbraket{\j_i(\cu)-\j_i(\cu')}{\j_i(\cu)-\j_i(\cu')}\ ,
\end{equation}
where we have neglected subleading terms in $1/\g$.
The accumulated discrepancy
$\sbraket{\j_i(\cu)-\j_i(\cu')}{\j_i(\cu)-\j_i(\cu')}$
is a function of various simulation parameters (in particular
the total MD time $\t_{MD}$), but it is  independent of $\g$.
On the final configuration we would thus have $S_{ker}\to\infty$
for $\g\to\infty$, implying that every trial configuration
would be rejected.
Combining both requirements, we must therefore choose $\g$ of order one.

There are three more parameters to discuss:
the number of ``lifted'' eigenvalues $n$,
the total MD time per trajectory $\tau_{MD}$,
and the MD time increment $\d\tau$.
While existing numerical simulations of the super-critical $\det(H_W^2)$
using the ordinary HMC algorithm
are confined to the sector with topological charge zero,
the results can guide us
in the choice of the remaining parameters.
Our conclusion is that we advocate first trying
THMC with values of $\tau_{MD}$ and $\d\tau$ similar to those used
in Refs.~\cite{pv,jlqcd,jlqcdfig}, and with a very small value of $n$,
perhaps even $n=1$.

The first interesting lesson is that, even at
rather large lattice spacings (with $a^{-1}$ as low as 1.5~GeV),
near-zero eigenvalues of $H_W$ were efficiently suppressed by the inclusion of
$\det(H_W^2)$ in the Boltzmann weight.
What this means is that, even if we were to choose an unrealistically large
value for $n$ (\eg, $n=100$), topology change during a given MD trajectory
should still take place mostly through the zero crossing of
just one, or very few, eigenvalues.  In principle, then, THMC will
restore ergodicity even if we allow for just one crossing
per trajectory, which, in turn, is made possible
by choosing any non-zero value for $n$. Because multiple crossings
per trajectory should  be much less frequent,
a plausible strategy is to make no special effort to allow for them.

The next question is about the effect of
the number $n$ of eigenmodes kept out of the MD evolution on
the acceptance rate.  As explained in Sec.~\ref{basic},
when using the THMC algorithm the acceptance rate will
be less than one even in the limit $\d\t\to 0$ where
$\ch_{MD}(\p',\cu')=\ch_{MD}(\p,\cu)$.
In the deterministic implementation of Sec.~\ref{determine},
acceptance is controlled by the ratio $\det(\ca(\cu'))/\det(\ca(\cu))$
in the limit $\d\t\to 0$.
Based on the model of App.~\ref{acc} we have concluded that the acceptance
rate is likely to go down with increasing $n$,
and thus it is desirable to keep $n$ as small as possible.

In the stochastic implementation, the metropolis test involves
the additional factor
  $\d S_{ker} = S_{ker}(\cu') - S_{ker}(\cu)$  (\cf~Eq.~(\ref{DWFP})).
Let us discuss the role of $\d S_{ker}$ at a qualitative level,
considering first the case $n=1$.
Because $\g=O(1)$, we clearly have $S_{ker}(\cu)=O(1)$ at the beginning
of the trajectory.  During the MD evolution the shape and location
of the (localized) eigenmode $\sket{\j_1}$ changes by an unknown
amount, and at the end of the trajectory $S_{ker}(\cu')$ will
assume another $O(1)$ value.
If we would now allow for larger values of $n$, the size of the interval
containing the most probable values of $\d S_{ker}$ will increase.
Positive values of $\d S_{ker}$,
which one expects to occur a fraction of the time,
will thus tend to further decrease the acceptance rate with increasing $n$.
We conclude that one should indeed try to keep $n$ as small as possible.

This brings us to the  trade-off between tunneling and acceptance.
While it remains
to be tested, we think that it is not implausible that THMC with
values as small as $n=1$ or $n=2$ will produce an appreciable tunneling rate.
Many of the results of Refs.~\cite{pv,jlqcd} are relevant for this
question, but particularly illuminating is Fig.~2 of Ref.~\cite{jlqcdfig}
which reports the results of an experimental two-flavor simulation.
We consider the left panel, generated with $\det(H_W^2)$ in
the Boltzmann weight.\footnote{
  We note that Refs.~\cite{jlqcd,jlqcdfig} actually used
  $\det(H_W^2)/\det(H_W^2+\mu^2)$ with $a\m=O(1)$, in order to reduce the
  contribution of the large eigenvalues of $H_W^2$ to the fermion force.
}
This plot shows the lowest few eigenvalues of $H_W$ as a function
of MD time. The MD time interval corresponding to one trajectory
was $\t_{MD}=0.5$, with $\t_{MD}/\d\t=O(20)$.
We learn from the plot that, frequently, $H_W$ has a small eigenvalue
that changes by nearly 100\% over a one-trajectory interval.
This is an encouraging result, suggesting that if one would switch from HMC
to THMC, there is a good change that zero crossings will take place.

Two worries immediately come to mind.
Consider those trajectories where one eigenvalue has changed
by a relatively large amount and became exceptionally small
toward the end of the trajectory.  As already noted, such trajectories
are relatively frequent in the data.
However, it is often true that that particular eigenvalue
is not the smallest, nor even the next-smallest (in absolute value),
at the beginning of the trajectory.  If we would now re-run the
trajectory from the same starting point with THMC, but use only $n=1$
or $n=2$ and no higher value, our selection of the initial lowest eigenvalue(s)
would miss that particular eigenvalue,
which is the best candidate for crossing.  In such a case, THMC would not do its job,
because the $\sket{\c_i}$ would likely not have much overlap with the
eigenvector corresponding to this eigenvalue, \cf\ Eq.~(\ref{I}).
Nevertheless, for some fraction of the configurations
the best-candidate eigenvalue did actually
start as the smallest one.  Thus, even $n=1$ should produce
a non-vanishing tunneling rate.

A related worry is that when an eigenvalue became exceptionally small
at the end of the trajectory,
often the final configuration was rejected.\footnote{
  This can be seen from the discontinuities in the spectral flows
  at integer values of the horizontal axis (``traj''):
  when a configuration was rejected the next trajectory was started
  with the same initial configuration as the previous one.
}
Once again, part of the answer is that some configurations were accepted
even though they had one exceptionally small eigenvalue.
If we would switch to THMC for the next trajectory, that eigenvalue would
now be selected even with $n=1$, and would have an appreciable
chance to cross zero.

However, more importantly, with ordinary HMC,
any eigenvalue must eventually be deflected away from zero because of
the unboundedly growing fermion force.
In contrast, with THMC any large repelling force originating
from one of the $n$ eigenvalues which were the lowest ones
at the beginning of the trajectory is eliminated.
An eigenvalue motion that, with standard HMC, had
been slowed down by the unboundedly growing repelling force,
will speed up under THMC.  Therefore,
we expect that many of the eigenvalue flows that ended up very close to
zero (but with, invariably, no crossing) when HMC was used,
would move more rapidly when we switch to THMC.   In fact, they may well cross
zero during the same MD time interval, and end up with the opposite sign and
well away from being exceptionally small by the end of the trajectory, thus
increasing considerably the acceptance probability of the configuration
at the end of the trajectory.

Finally, we consider how the location of an initially low-lying mode
evolves. If the eigenmode would ``drift'' appreciably away from
its original location, this would give rise to increased $S_{ker}$
on the final configuration, and once again decrease the acceptance rate.
Worse, the overlap between the $\sket{\c_i}$ and this mode would
rapidly decrease, thus possibly decreasing the smallest eigenvalues
of $\cm$, and hence re-introducing barriers in the MD evolution.
This, however, is unlikely to happen for the following reason.

Having an exceptionally low eigenvalue requires two things:
values for some plaquettes substantially different from one,
as well as a certain ``fine tuning''
of the link variables relative to each other.
This picture, put forward in  Ref.~\cite{bnn}, is corroborated
by the fact that large plaquette values are far more frequent
compared to exceptionally low eigenvalues.  Now, because
the momenta conjugate to each link variable are basically
uncorrelated with each other, the likely outcome of the MD evolution
is that any finely-tuned relation
between local plaquettes will be quickly destroyed.
Once again this expectation is supported by Fig.~2 of Ref.~\cite{jlqcdfig},
which shows that exceptionally-low eigenvalues tend to evolve much more rapidly
compared to the ``bulk'' of (not-so-small) eigenvalues.
This supports the conjecture that indeed
the necessary fine-tuned dislocations only exist for a relatively
short MD time, in agreement with the
general observation that dislocations supporting exceptionally low
eigenvalues are rare.  In contrast, a ``drift'' of the
location of the eigenmode (in which the eigenvalue is kept very small)
would require an orchestrated motion
of the dislocation with almost no change in its shape, something
which is very unlikely to happen.

The picture that emerges, then, is that of a localized eigenvector
that remains more or less in place and localized, with a rapidly changing
eigenvalue and, presumably, correspondingly rapid changes
in its detailed shape.  This implies that $\langle\psi_i|\chi_i\rangle$
remains of order one, and thus no MD barrier is re-generated.
We emphasize, as above, that  this picture
only needs to hold for a sizable fraction of the  low-lying eigenmodes
that exist at the start of the MD evolution trajectories.

%%%%%%%
%\newpage
\section{\label{disc} Conclusion}
%%%%%%%
In this paper we proposed a new variant of the HMC algorithm.
The aim of the \textit{tunneling} HMC algorithm
is to allow for eigenvalue crossings through zero even in situations where
the  determinant containing such eigenvalues
is included in the Boltzmann weight.
The application we considered explicitly is lattice QCD
with domain-wall or overlap fermions
and with the auxiliary determinant of $H_W^2$.  The THMC algorithm may
find other applications as well.\footnote{
  For example, we envisage using THMC as an ingredient in the simulation of
  the ghost sector of equivariantly gauge-fixed Yang-Mills
  lattice theories \cite{ebrst}.
}

A major issue is the competition between  tunneling rate and acceptance.
We have discussed in some detail why we believe that
appreciable tunneling is not incompatible with reasonably high acceptance.
Setting aside the details, the underlying reason why this can be true
is that the difference between the guiding and metropolis
hamiltonians, $\ch-\ch_{MD}$, is not an extensive quantity.
For any lattice volume, in order to allow for tunneling between
topological sectors, we only need to keep a fixed, and very small,
number of degrees of freedom outside of the guiding hamiltonian.
Once ergodicity is restored via THMC, the correct Boltzmann weight
will automatically be reproduced, in all topological sectors.
Therefore, the cost of THMC does not grow with volume.  We stress that
our semi-quantitative conclusions will have to be tested numerically.

We discussed two concrete implementations of THMC, a deterministic and a
stochastic one.
We believe that the stochastic implementation of THMC is  likely to be
of most practical interest (especially for domain-wall simulations),
because of its smaller computational overhead.  For this to be the case,
any further decrease in acceptance coming from the extra element
present in the stochastic implementation, $S_{ker}$, should not be too
large.

In comparison to standard HMC, the extra cost of THMC comes
partly from its more elaborate structure.  For example,
it takes more operations to perform
a single multiplication of a vector by $\cm$,
as compared to $H_W^2$.  However, it is likely that the major extra
cost of THMC will come from reduced acceptance: if acceptance is
a factor of two smaller, this immediately translates into a factor
of two extra cost.  Any extra cost of THMC will have to be weighed
against the physical merits of the generated ensemble.
For example, the fixed-topology simulations of JLQCD \cite{jlqcd}
give rise to enhanced finite-volume effects proportional
to $1/(m_\p^2 V)$ \cite{bcnw}.  Ultimately, the only way to keep
such enhanced finite-size effects under systematic
control is to simulate bigger lattice volumes,
something that carries its own price tag.  Clearly, it is the cost for
``equal-quality'' physics output which must be compared.
For domain-wall fermion simulations, the extra cost of THMC must be
balanced against the improvement of chiral symmetry.

Are there possible alternatives to THMC?  The answer is yes.
A simple approach is to replace $\det(H_W^2)$
by $\det(H_W^2+\e^2)$ as the auxiliary determinant, where presumably
$\e^2$ would be chosen comparable to a typical small eigenvalue of $H_W^2$,
much like the parameter $\a$ introduced in Eq.~(\ref{RG}).  Because
there are no exact zero modes any more, one can simulate $\det(H_W^2+\e^2)$
using ordinary HMC.\footnote{
  We learned about this alternative approach from Norman Christ.
}
In comparison with THMC, this avoids altogether
any drop in acceptance associated with a non-zero $\ch-\ch_{MD}$.
At the same time,
if $\e$ is chosen small enough, the low-lying eigenvalues of $H_W^2$
would still be suppressed.  An additional benefit of simulating $\det(H_W^2+\e^2)$
is that no new code needs to be written, and there are fewer new parameters
that one has to experiment with for optimal performance.

In spite of the obvious advantages of this alternative,
we believe that
it is an open question which solution is better. When using
$\det(H_W^2+\e^2)$ one will have to compromise the value of $\e$ between
two conflicting requirements: the basic need to suppress the low-lying
eigenvalues, and the need not to suppress them too much
so as to allow for tunneling.  In contrast, the advantage
of THMC is that it solves one problem at a time:
the  reduction of an abundance of  low-lying eigenvalues
comes from $\det(H_W^2)$, while the restoration of tunneling
comes from the modification of the algorithm.

While in this paper we have assumed that the auxiliary determinant
is $\det(H_W^2)$, the THMC algorithm evidently generalizes to other
even powers as this determinant.  If the auxiliary determinant
is $\det(H_W^{2k})$, the near-zero spectral density will behave
like $\r(\l) \sim \l^{2k}$.  In a domain-wall simulation,
the residual mass will thus scale like $1/L_5^{2k+1}$.
This constitutes a bigger pay-back for any increase of $L_5$ in
the region where the residual mass is dominated by the near-zero
modes.

The choice of the JLQCD collaboration to run fixed-topology
dynamical overlap simulations stems from several technical advantages
arising from the introduction of $\det(H_W^2)$.
The (much!) smaller near-zero spectral density
leads to a smaller range for $1/|H_W|$, and an affordable cost
for an adequate approximation of the sign function $H_W/|H_W|$.
The costly reflection/refraction step is not needed by design,
because no eigenvalue of $H_W$ ever reaches zero during the MD evolution.
With THMC, the same reduction in the near-zero spectral density
is achieved, with all the ensuing advantages.  Of course,
reflection/refraction steps cannot be avoided if we want topology
to change.\footnote{
  See, however, Sec.~3.3 of Ref.~\cite{sf}.
}
But the number of attempted topology changes will remain small,
and will not grow with the volume, because it is controlled by
the number $n$ of eigenvalues which are relegated to $\det(\ca)$.
As we have argued above, this number can be chosen to be very small.
The cost of reflection/refraction steps should thus be tolerable.

Some of the difficulties in achieving a satisfactory topology-changing rate
are inherent.  While too-many small eigenvalues of $H_W$ is the problem
at relatively large lattice spacing, when the lattice spacing gets smaller
eventually the situation changes, because the
near-zero mode density decreases rapidly with the bare coupling.
This means that tunneling becomes rare, even without the auxiliary
determinant in the Boltzmann weight.
For other difficulties associated with tunneling from the $Q=0$
sector to sectors with a higher topological charge, see \eg~Ref.~\cite{sf}.
It is an open question
whether THMC can help in creating a bigger interval of lattice-spacing values
where the tunneling rate is satisfactory.

%%%%%%%%%%%%
%\newpage
\vspace{3ex}
\noindent {\bf Acknowledgments}
\vspace{3ex}

We would like to thank Tom DeGrand, Shoji Hashimoto, Urs Heller,
Tetsuya Onogi, Bob Sugar and
Pavlos Vranas for discussions, and Tom Blum and Amarjit Soni for
organizing the workshop ``Domain-Wall fermions at 10 Years," at which
many of these discussions took place.  We thank Shohi Hashimoto in particular
for detailed explanations of the work reported in Refs.~\cite{jlqcd,jlqcdfig}.
We thank the referee for catching some errors in the original manuscript.
YS thanks the Department of
Physics and Astronomy of San Francisco State University for hospitality.
MG was supported in part by the US Department of Energy.
YS was
supported by the Israel Science Foundation under grant no.~173/05.

\appendix

%%%%%%%
%\newpage
\section{\label{appA} Proof of Eq.~(\ref{RG})}
%%%%%%%
We represent $\det(H_W^2)$ as a Grassmann integral over $\j$ and $\bj$,
multiply the integrand by unity written as a new Grassmann integral
over $\J$ and $\bJ$, and perform the integration first
over $\j$ and $\bj$ and then over $\J$ and $\bJ$.
Keeping all normalization factors the result is
\begin{eqnarray}
  Z \;\equiv\; \det(H_W^2)  &=& \int \cd\j \cd\bj\; \exp(-\bj H_W^2 \j)
\label{block}
\\
  && \hspace{-10ex} = \;
  \a^{-n}\, \int \cd\j \cd\bj\; \prod_{i=1}^n d\J_i d\bJ_i\; \exp\left(
  -\bj H_W^2  \j
  - \a(\bJ - \bj Q^\dagger)(\J - Q \j) \right)
\NON
  && \hspace{-10ex} = \; \a^{-n}\; \det(\cm) \int \prod_{i=1}^n d\J_i d\bJ_i\;
  \exp(-\bJ \tca \J)
\NON
  && \hspace{-10ex} = \; \a^{-n}\; \det(\cm)\, \det(\tca)\ ,
\nonumber
\end{eqnarray}
where $\cm$ is given by Eq.~(\ref{RGb}) and
\begin{equation}
  \tca = \a - \a^2 Q \cm^{-1} Q^\dagger \ .
\label{A'}
\end{equation}
We next consider the two-point function
\begin{eqnarray}
  \tca^{-1}_{ij} =
  \svev{\J_i\,\bJ_j}
  &=& Z^{-1}\, \a^{-n}\,
  \int \cd\j \cd\bj\; \prod_{i=1}^n d\J_i d\bJ_i\;
\label{AA'}
\\
  && \times \exp\left(-\bj H_W^2  \j
  - \a(\bJ - \bj Q^\dagger)(\J - Q \j) \right)\; \J_i\,\bJ_j
\NON
  &=& \rule{0ex}{3.5ex}
  Z^{-1}\, \int \cd\j \cd\bj\; \exp(-\bj H_W^2 \j)\;
  \left( \a^{-1} \d_{ij} + (Q \j)_i\,(\bj Q^\dagger)_j\right) \ .
\nonumber
\end{eqnarray}
Performing the integral over $\j$ and $\bj$ we arrive at
Eq.~(\ref{RGc}) with the identification $\tca=\ca$.

%%%%%%%
%\newpage
\section{\label{acc} Acceptance rate in the deterministic implementation}
%%%%%%%
In this appendix we discuss a simple model for the decrease in
acceptance resulting from the omission of $\det(\ca)$ from the MD evolution,
within the deterministic implementation of THMC.
We will first consider the case $n=1$, namely,
a single eigenmode is omitted from the MD evolution.
We begin by writing down a general expression for the acceptance rate.
At the beginning of a trajectory,
$x=\det(\ca(\cu))$ follows a probability distribution $P(x)$.
At the end of the trajectory, $y=\det(\ca(\cu'))$ follows another
distribution $\hat{P}(y|x)$.  This distribution is conditional on $x$,
because the initial and final gauge field configurations
are not uncorrelated.  Ignoring step-size errors, \ie, assuming that
$\ch_{MD}(\pi,\cu)=\ch_{MD}(\pi',\cu')$,
Eq.~(\ref{ARmod}) gives an acceptance rate
\begin{equation}
\label{ARmodel}
P_{accept}=\int_0^1 dx\;P(x)\;\int_0^1 dy\;\hat{P}(y|x)\;{\rm min}\{1,y/x\}\ ,
\end{equation}
where have we rescaled the possible values of $\det(\ca)$ to
a unit interval.

In order to estimate $P_{accept}$ we need to make several assumptions.
The initial configuration $\cu$ is an equilibrated configuration
of the ensemble generated by the Boltzmann weight
$\exp(-S_g(\cu))\;\prod_i \l_i^2$, {\it cf.}~Eq.~(\ref{Z}).
We assume that the eigenvalue $\l_0$ with smallest absolute value
is localized, and practically uncorrelated with the remaining eigenvalues.
This suggests a probability distribution for this eigenvalue
which is proportional to $\l_0^2$ itself, for $\l_0^2 \ll 1$.
As for the support, we postulate a sharp-cutoff model, namely,
we assume that the probability distribution for the smallest eigenvalue
vanishes outside the interval $[0,C\a]$; inside this interval,
it increases with $\l_0^2$, starting from zero at $\l_0^2=0$.
Physically, $C\a$ is the average of the second-smallest eigenvalue
(squared).  Our choice of $\a$ (see Sec.~\ref{basic}) ensures that $C=O(1)$.
Treating $C$ as a free parameter thus allows us to get an idea on how
sensitively the drop in acceptance depends on the details of the theory.

What enters Eq.~(\ref{ARmodel}) is not directly the probability distribution for
$\l_0^2$ but, rather, that for the related quantity $x=\det(\ca(\cu))$.
Treating $x$ as the independent variable, we specify the probability
distribution by postulating $P(x) \propto \l_0^2(x) = x\a/(\a-x)$ over its
support, where the last equality follows from Eq.~(\ref{AMb}).
Similarly using Eq.~(\ref{AMb}) to obtain the $x$-range
that corresponds to the previously selected range of $\l_0^2$,
followed by a rescaling to the unit interval,
we obtain the probability distribution\footnote{
  See below for more discussion of this crude approximation.
}
\begin{equation}
  P(x) = N \frac{x}{C+1-Cx}\ ,   \qquad 0\le x\le 1\ ,
\label{Px}
\end{equation}
where
\begin{equation}
  \frac{1}{N} = \frac{C+1}{C^2} \log(C+1) - \frac{1}{C} \ .
\label{N}
\end{equation}

We next turn to the conditional probability $\hat{P}(y|x)$.
The small eigenvalues of $H_W^2$ on a typical
configuration are expected to be well below its mobility edge,
and thus to be localized with a range of order $a$ \cite{gs,gss}.
These eigenvalues are only sensitive to the details of the gauge
field near the corresponding localized modes, and these gauge fields
fluctuate on a short (MD) time scale.
It is therefore not unreasonable to expect that, for typical values
of the MD time interval $\tau_{MD}$ for one complete trajectory,
the variable $y$ actually becomes decorrelated from $x$,
$\hat{P}(y|x)=\hat{P}(y)$.  Moreover,
the THMC algorithm is designed to remove the lowest eigenvalue
from the MD force.  This suggests that the probability distribution
$\hat{P}(y)$ is rather flat.  For the sake of simplicity
we thus take $\hat{P}(y)$ to be constant, and have the same
support as $P(x)$, which amounts to choosing $\hat{P}(y)=1$ for $0\le y \le 1$
(after a similar rescaling).
Finally, plugging this into Eq.~(\ref{ARmodel})
and using Eq.~(\ref{Px}) we obtain
\begin{equation}
  P_{accept} = \frac{1}{2} + \frac{N-2}{4C}\ .
\label{Pvalue}
\end{equation}
The limits $C\to 0$ and $C\to\infty$ give rise to
$P_{accept}=2/3$ and $P_{accept}=1/2$, respectively.

Let us now generalize this model to the case that two eigenmodes are omitted
from the MD evolution.  We write $\det(\ca(\cu)) = x_1 x_2$,
where $x_1$ and $x_2$ are related to the corresponding
eigenvalues in the same way that $x$ was previously related to
the smallest eigenvalue. These modes are expected to be localized,
and uncorrelated, at the beginning of the trajectory.
We may thus assume the same probability distribution
as before, separately for $x_1$ and for $x_2$.\footnote{
  Here we assume that the smallest and second-smallest eigenvalues
  are associated to $x_1$ or to $x_2$ based not on their magnitude
  but rather on some other criterion such as, for example, the location
  of the eigenmodes.  This justifies using the same probability
  distribution for $x_1$ and $x_2$, because $x_2$ will correspond
  to the smallest eigenvalue just as often as $x_1$.
}
Similarly, writing $\det(\ca(\cu')) = y_1 y_2$,
we will assume the same flat distribution for $y_1$ and $y_2$
at the end of the trajectory prior
to the metropolis test.
The acceptance rate is then given by
\begin{equation}
\label{ARmodel2}
P_{accept}=\int_0^1 dx_1 dx_2\;P(x_1)P(x_2)\;
\int_0^1 dy_1 dy_2\;{\rm min}\{1,(y_1 y_2)/(x_1 x_2)\}\ ,
\end{equation}
where $P(x)$ is given by Eq.~(\ref{Px}).
Varying $C$ in the range of 0.01 to 100 we now find that $P_{accept}$
varies between roughly 50\% and 35\% respectively.

Returning to the case that just one eigenmode is omitted from the MD
evolution, our model, the probability distribution
$P(x)$ of Eq.~(\ref{Px}), neglects most factors in the Boltzmann weight.
We have ignored the pure-gauge action, as well as all the remaining
eigenvalues of $H_W$.
In addition, we neglected the jacobian that arises if
we treat $x=\det(\ca(\cu))$
as an independent integration variable.  The motivation for
neglecting all these factors
is twofold.  First, clearly, it would be difficult to incorporate
all factors  accurately.  Also, while all the neglected factors are unlikely to
be constant for the range where our chosen $P(x)$ is non-zero,
they are also not expected to vanish.  The most crucial feature of
the exact, yet unknown, probability distribution is that it vanishes with
$\l_0^2$.  This essential feature is captured by our model.
Of course, many other models with the same behavior for $\l_0^2\ll 1$
could be conceived.  The free parameter
$C$ reflects to some extent the arbitrariness of the model.
%$C$ mimics to some extent the effect of the neglected factors,
%or equivalently, compensates for the arbitrariness of the model.
The fact that rather large changes in the value of $C$ do not lead
to a dramatic change in the resulting acceptance is, we believe,
an encouraging sign that the acceptance rate will indeed be roughly
in the range predicted by our model.  Of course, ultimately the only way
to verify these expectations is through numerical tests of the THMC algorithm.

%%%%%%%%%%%%%%%%%%%%%%%%%%%
%\newpage


\begin{thebibliography}{99}

\bibi{kaplan}
  D.~B.~Kaplan,
  %``A Method for simulating chiral fermions on the lattice,''
  Phys.\ Lett.\  B {\bf 288}, 342 (1992)
  [arXiv:hep-lat/9206013].
  %%CITATION = PHLTA,B288,342;%%

\bibi{ys}
  Y.~Shamir,
  %``Chiral fermions from lattice boundaries,''
  Nucl.\ Phys.\  B {\bf 406}, 90 (1993)
  [arXiv:hep-lat/9303005];
  %%CITATION = NUPHA,B406,90;%%
  V.~Furman and Y.~Shamir,
  %``Axial Symmetries In Lattice QCD With Kaplan Fermions,''
  Nucl.\ Phys.\  B {\bf 439}, 54 (1995)
  [arXiv:hep-lat/9405004].
  %%CITATION = NUPHA,B439,54;%%

\bibi{mresCPPACS}
  A.~Ali Khan {\it et al.}  [CP-PACS Collaboration],
  %``Chiral properties of domain-wall quarks in quenched QCD,''
  Phys.\ Rev.\ D {\bf 63}, 114504 (2001)
  [arXiv:hep-lat/0007014];
  %%CITATION = HEP-LAT 0007014;%%
  %``Eigenvalues of the hermitian Wilson-Dirac operator and chiral  properties
  %of the domain-wall fermion,''
  Nucl.\ Phys.\ Proc.\ Suppl.\  {\bf 94}, 725 (2001)
  [arXiv:hep-lat/0011032].
  %%CITATION = HEP-LAT 0011032;%%

\bibi{mresRBC}
  T.~Blum {\it et al.} [RBC Collaboration],
  %``Quenched lattice QCD with domain wall fermions and the chiral limit,''
  Phys.\ Rev.\ D {\bf 69}, 074502 (2004)
  [arXiv:hep-lat/0007038];
  %%CITATION = HEP-LAT 0007038;%%
  K.~Orginos  [RBC Collaboration],
  %``Chiral properties of domain wall fermions with improved gauge actions,''
  Nucl.\ Phys.\ Proc.\ Suppl.\  {\bf 106}, 721 (2002)
  [arXiv:hep-lat/0110074];
  %%CITATION = HEP-LAT 0110074;%%
  Y.~Aoki {\it et al.} [RBC Collaboration],
  %``Domain wall fermions with improved gauge actions,''
  Phys.\ Rev.\ D {\bf 69}, 074504 (2004)
  [arXiv:hep-lat/0211023].
  %%CITATION = HEP-LAT 0211023;%%

\bibi{yspt}
  Y.~Shamir,
  %``New domain-wall fermion actions,''
  Phys.\ Rev.\  D {\bf 62}, 054513 (2000)
  [arXiv:hep-lat/0003024].
  %%CITATION = PHRVA,D62,054513;%%

\bibi{pvsm}
  P.~M.~Vranas,
  %``Chiral symmetry restoration in the Schwinger model with domain wall
  %fermions,''
  Phys.\ Rev.\  D {\bf 57}, 1415 (1998)
  [arXiv:hep-lat/9705023].
  %%CITATION = PHRVA,D57,1415;%%

\bibi{AT}
  S.~Aoki and Y.~Taniguchi,
  %``One loop calculation in lattice QCD with domain-wall quarks,''
  Phys.\ Rev.\  D {\bf 59}, 054510 (1999)
  [arXiv:hep-lat/9711004].
  %%CITATION = PHRVA,D59,054510;

\bibi{scri}
  R.~G.~Edwards, U.~M.~Heller and R.~Narayanan,
  %``Spectral flow, chiral condensate and topology in lattice QCD,''
  Nucl.\ Phys.\  B {\bf 535}, 403 (1998)
  [arXiv:hep-lat/9802016];
  %%CITATION = NUPHA,B535,403;%%
    R.~G.~Edwards, U.~M.~Heller and R.~Narayanan,
  %``A study of practical implementations of the overlap-Dirac operator in  four
  %dimensions,''
  Nucl.\ Phys.\  B {\bf 540}, 457 (1999)
  [arXiv:hep-lat/9807017].
  %%CITATION = NUPHA,B540,457;%%

\bibi{bnn}
  F.~Berruto, R.~Narayanan and H.~Neuberger,
  %``Exact local fermionic zero modes,''
  Phys.\ Lett.\  B {\bf 489}, 243 (2000)
  [arXiv:hep-lat/0006030].
  %%CITATION = PHLTA,B489,243;%%

\bibi{gs}
  M.~Golterman and Y.~Shamir,
  %``Localization in lattice QCD,''
  Phys.\ Rev.\  D {\bf 68}, 074501 (2003)
  [arXiv:hep-lat/0306002].
  %%CITATION = PHRVA,D68,074501;%%

\bibi{hn}
  H.~Neuberger,
  %``Exactly massless quarks on the lattice,''
  Phys.\ Lett.\  B {\bf 417}, 141 (1998)
  [arXiv:hep-lat/9707022].
  %%CITATION = PHLTA,B417,141;%%

\bibi{pv}
  P.~M.~Vranas,
  %``Domain wall fermions in vector theories,''
  arXiv:hep-lat/0001006;
  %%CITATION = HEP-LAT/0001006;%%
  P.~M.~Vranas,
  %``Gap domain wall fermions,''
  Phys.\ Rev.\  D {\bf 74}, 034512 (2006)
  [arXiv:hep-lat/0606014].
  %%CITATION = PHRVA,D74,034512;%%

\bibi{jlqcd}
%\bibitem{Fukaya:2006vs}
  H.~Fukaya, S.~Hashimoto, K.~I.~Ishikawa, T.~Kaneko, H.~Matsufuru, T.~Onogi
  and N.~Yamada [JLQCD Collaboration],
  %``Lattice gauge action suppressing near-zero modes of H(W),''
  Phys.\ Rev.\  D {\bf 74}, 094505 (2006)
  [arXiv:hep-lat/0607020].
  %%CITATION = PHRVA,D74,094505;%%

\bibi{topo}
  R.~Narayanan and H.~Neuberger,
  %``Chiral determinant as an overlap of two vacua,''
  Nucl.\ Phys.\  B {\bf 412}, 574 (1994)
  [arXiv:hep-lat/9307006].
  %%CITATION = NUPHA,B412,574;%%

\bibi{hmc}
  S.~Duane, A.~D.~Kennedy, B.~J.~Pendleton and D.~Roweth,
  %``HYBRID MONTE CARLO,''
  Phys.\ Lett.\  B {\bf 195} (1987) 216.
  %%CITATION = PHLTA,B195,216;%%

\bibi{bcnw}
  R.~Brower, S.~Chandrasekharan, J.~W.~Negele and U.~J.~Wiese,
  %``QCD at fixed topology,''
  Phys.\ Lett.\  B {\bf 560}, 64 (2003)
  [arXiv:hep-lat/0302005];
  %%CITATION = PHLTA,B560,64;%%
  T.~Onogi, talk at the workshop ``Domain-wall fermions at 10 years,''
  {\tt https://www.bnl.gov/riken/dwf/}\ ,
  S.~Aoki, H.~Fukaya, S.~Hashimoto and T.~Onogi, in progress.

\bibi{rev}
%\bibi{Kennedy:2006ax}
  A.~D.~Kennedy,
  %``Algorithms for dynamical fermions,''
  arXiv:hep-lat/0607038;
  %%CITATION = HEP-LAT 0607038;%%
%\bibi{Clark:2006wq}
  M.~A.~Clark,
  %``The rational hybrid Monte Carlo algorithm,''
  arXiv:hep-lat/0610048.
  %%CITATION = HEP-LAT/0610048;%%

\bibi{dk}
  S.~Duane and J.~B.~Kogut,
  %``THE THEORY OF HYBRID STOCHASTIC ALGORITHMS,''
  Nucl.\ Phys.\  B {\bf 275}, 398 (1986).
  %%CITATION = NUPHA,B275,398;%%

\bibi{sdg}
  T.~A.~DeGrand and S.~Schaefer,
  %``Physics issues in simulations with dynamical overlap fermions,''
  Phys.\ Rev.\  D {\bf 71}, 034507 (2005)
  [arXiv:hep-lat/0412005].
  %%CITATION = PHRVA,D71,034507;%%

\bibi{NC}
  N.~Cundy,
  %``Small Wilson Dirac operator eigenvector mixing in dynamical overlap
  %hybrid Monte-Carlo,''
  arXiv:0706.1971 [hep-lat].
  %%CITATION = ARXIV:0706.1971;%%

\bibi{rflr}
%\bibitem{Fodor:2003bh}
  Z.~Fodor, S.~D.~Katz and K.~K.~Szabo,
  %``Dynamical overlap fermions, results with hybrid Monte-Carlo algorithm,''
  JHEP {\bf 0408}, 003 (2004)
  [arXiv:hep-lat/0311010].
  %%CITATION = JHEPA,0408,003;%%

\bibi{sf}
  S.~Schaefer,
  %``Algorithms for dynamical overlap fermions,''
  PoS {\bf LAT2006}, 020 (2006)
  [arXiv:hep-lat/0609063].
  %%CITATION = POSCI,LAT2006,020;%%

\bibi{jlqcdfig}
%\bibitem{Matsufuru:2006xr}
  H.~Matsufuru {\it et al.}  [JLQCD Collaboration],
  %``Improvement of algorithms for dynamical overlap fermions,''
  PoS {\bf LAT2006}, 031 (2006)
  [arXiv:hep-lat/0610026].
  %%CITATION = POSCI,LAT2006,031;%%

\bibi{ebrst}
  M.~Golterman and Y.~Shamir,
  %``SU(N) chiral gauge theories on the lattice,''
  Phys.\ Rev.\  D {\bf 70}, 094506 (2004)
  [arXiv:hep-lat/0404011];
  %%CITATION = PHRVA,D70,094506;%%
 %``Running couplings in equivariantly gauge-fixed SU(N) Yang-Mills theories,''
  Phys.\ Rev.\  D {\bf 73}, 014510 (2006)
  [arXiv:hep-lat/0511042].
  %%CITATION = PHRVA,D73,014510;%%

\bibi{gss}
  M.~Golterman, Y.~Shamir and B.~Svetitsky,
  %``Mobility edge in lattice QCD,''
  Phys.\ Rev.\  D {\bf 71}, 071502 (2005)
  [arXiv:hep-lat/0407021];
  %%CITATION = PHRVA,D71,071502;%%
  %``Localization properties of lattice fermions with plaquette and improved
  %gauge actions,''
  Phys.\ Rev.\  D {\bf 72}, 034501 (2005)
  [arXiv:hep-lat/0503037].
  %%CITATION = PHRVA,D72,034501;%%

\end{thebibliography}
\end{document}